# Ultrafast and selective coherent population transfer in four-level atoms by a single frequency chirped few-cycle pulse

Parvendra Kumar and Amarendra K. Sarma*

*Department of Physics, Indian Institute of Technology Guwahati, Guwahati-781039, Assam, India.*
*Corresponding author: aksarma@iitg.ernet.in*

We report and propose a simple scheme to achieve the ultrafast and selective population transfer in four-level atoms by utilizing a single frequency chirped few-cycle pulse. It is demonstrated that the almost complete population may be transferred to the preselected state of atoms just by manipulating the so called chirp offset parameter. The robustness of the scheme against the variation of laser pulse parameters is also investigated. The proposed scheme may also be useful for the selective population transfer in molecules. © 2013 Optical Society of America

Coherent population transfer (CPT) to a particular state of atoms and molecules is highly desirable for many practical applications such as quantum information processing, interferometry, optical control of chemical reactions and precise spectroscopy. Recently, there has been a resurgence of interest in the CPT due to the technological advancement in the generation of shaped femtosecond laser pulses [1–5]. Mainly three effective methods are known to transfer the complete population in a particular state; $\pi$-pulse [6], stimulated Raman adiabatic passage (STIRAP) [7-9], and adiabatic rapid passage (ARP) [10]. However, $\pi$-pulse scheme is highly sensitive to pulse parameter and resonant condition. In STIRAP scheme two photon resonance and time delay between two pulses are crucial for effective population transfer. The third scheme, ARP consists the sweeping of laser carrier frequency through the atomic or molecular resonance and seems to be most promising one with femtosecond pulses because the sweeping of frequency in femtosecond pulses could be done effectively owing to the large frequency bandwidth of femtosecond pulses. Recently, much attention has been paid towards the coherence creation and CPT by utilizing a single frequency chirped pulse in $\Lambda$-like three level and four level atoms [11, 12]. In particular, S. A. Malinovskaya et al. [12] demonstrate the CPT in ultra cold Rb atoms using a single linearly chirped pulse. In this work, we demonstrate, first time to the best of our knowledge, the selective and efficient CPT in Y-like four-level Na atoms by utilizing a single frequency chirped few-cycle pulse. It is shown that the selective CPT could be achieved by just manipulating the chirp offset parameter. The phenomena of CPT is investigated by numerically solving the appropriate density matrix equations beyond the rotating wave approximation as it has been pointed out by many authors that the so-called rotating wave approximation (RWA) do not hold when one deals with few-cycle pulses [13, 14]. In addition, we assume that all the atomic relaxation times are considerably longer than the interaction times. Our proposed scheme is depicted in Fig. 1.

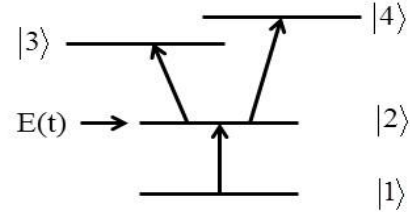

Fig.1 Schematic of the scheme

In Fig. 1, the levels, $|1\rangle, |2\rangle, |3\rangle$, and $|4\rangle$ represent the 3s, 3p, 5s, 4d states of sodium atoms respectively. The complete Hamiltonian without invoking RWA approximation, which describes the interaction of single pulse with four-level atoms, is given by:

$$\hat{H} = \hbar \begin{pmatrix} \omega_1 & -\Omega_R(t) & 0 & 0 \\ -\Omega_R(t) & \omega_2 & -\beta\Omega_R(t) & -\gamma\Omega_R(t) \\ 0 & -\beta\Omega_R(t) & \omega_3 & 0 \\ 0 & -\gamma\Omega_R(t) & 0 & \omega_4 \end{pmatrix}$$

Here $\Omega_R(t) = \mu_{12} E(t)/\hbar$, $\mu_{12}$ is the transition dipole moment of $|1\rangle \to |2\rangle$ transition. For the generalization of problem to other atoms as well, the other transition dipole moments; $\mu_{23}$, $\mu_{24}$ are chosen as follows: $\mu_{23} = \beta\mu_{12}$ and $\mu_{24} = \gamma\mu_{12}$. Here, $\beta$ and $\gamma$ are the coefficient of dipole moments. The electric field part of pulse is defined as follows: $E(t) = E_0 \exp\left(-(t/\tau_p)^2\right)\cos(\omega t + \delta(t))$. Here $E_0$ is the peak amplitude of pulse envelope, $\tau_{FWHM} = 1.177\tau_p$, $\omega$ is the central frequency and $\delta(t)$ is the time varying phase. In this work, we consider the same temporal profile of $\delta(t)$ as considered by J. J.

Carrera et al. [15] and defined as $\delta(t) = -\alpha \tanh\left(\frac{t+t_0}{\tau}\right)$. The chirped form of pulse may be controlled by manipulating the three parameters $\alpha$, $t_0$ and $\tau$. In this work, these three parameters are termed as frequency sweeping, chirp offset and chirp steepening parameters respectively. The instantaneous frequency of the pulse has the form: $\omega(t) = \omega - \frac{\alpha}{\tau}\text{sech}^2\left(\frac{t+t_0}{\tau}\right)$. We use the following density matrix formalism to study the population dynamics.

$$\frac{d\rho_{nm}}{dt} = \frac{-i}{\hbar}\left[\hat{H},\hat{\rho}\right]_{nm}$$

Here $\rho_{nm}(n, m = 1 \rightarrow 4)$ is the component of density matrix, $\rho_{nn}$ denotes the population of $n^{th}$ level while $\rho_{nm}$ denotes the coherence between level 'n' and level 'm'. We use the following parameters: $\omega_{21}$=3.19 rad/fs, $\omega_{32}$ = 3.06 rad/fs, $\omega_{42}$ =3.30 rad/fs, peak Rabi frequency, $\Omega_R(0)$ =0.60 rad/fs, $\tau_p$=16.5 fs, $t_0 = \pm 16.5$ fs, $\tau = 16.5$ fs, $\omega = 3.6 \text{rad/fs}$, $\beta$ =0.90, $\gamma$ =1.10 and $\alpha$ = 10.0 rad. It is worth to mention that the aforementioned pulse parameters are investigated in order to achieve the selective and maximum population transfer. Fig. 2 depicts the temporal evolution of pulse frequency, pulse envelope and populations in different states.

varying frequency is resonant with the frequency of $|2\rangle \rightarrow |3\rangle$ transition at t ≈ -18 fs and at later time t ≈ -8 fs is resonant with the frequency of $|1\rangle \rightarrow |2\rangle$ transition. This counterintuitive sequence makes $|2\rangle \rightarrow |4\rangle$ transition nearly forbidden and leads to the almost complete (98.40 %) population transfer to state $|3\rangle$ as could be observed from Fig. 2(b). On the other hand, it might be clear from Fig. 2 (c) that the pulse is interacting with $|1\rangle \rightarrow |2\rangle$ and $|2\rangle \rightarrow |4\rangle$ transitions in counterintuitive manner also because with chosen chirp offset parameter, $t_0$ = -16.5 fs, initially the time varying frequency is resonant with the frequency of $|2\rangle \rightarrow |4\rangle$ transition at t ≈ 2 fs and at later time t ≈ 6 fs is resonant with the frequency of $|1\rangle \rightarrow |2\rangle$ transition. This counterintuitive sequence makes the $|2\rangle \rightarrow |3\rangle$ transition nearly forbidden and leads to the almost complete (98.50 %) population transfer to state $|4\rangle$ as could be observed from Fig. 2(d). Hence the selective population transfer could be achieved by just manipulating the chirp offset parameter. It is important to verify the robustness of the scheme against the variation of the pulse parameters for practical realization of the scheme. So, in Fig. 3 we present the simulation result for the variation of the final population transfer to the state $|3\rangle$, i.e., $\rho_{33}(\infty)$ and state $|4\rangle$, i.e., $\rho_{44}(\infty)$ with frequency sweeping and chirp steeping parameters.

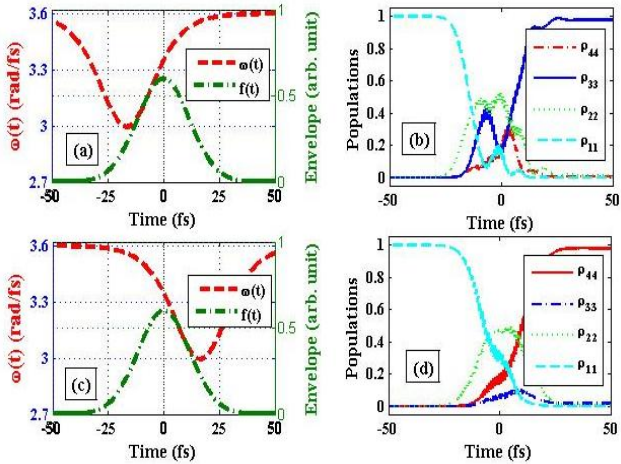

Fig. 2 (Color online) Temporal evolution of pulse frequency (a, c), pulse envelope (a, c) and populations (b, d).

It could be understood from Fig. 2(a) that the pulse is interacting with $|1\rangle \rightarrow |2\rangle$ and $|2\rangle \rightarrow |3\rangle$ transitions in counterintuitive manner because with chosen chirp offset parameter, $t_0$ = +16.5 fs, initially the time

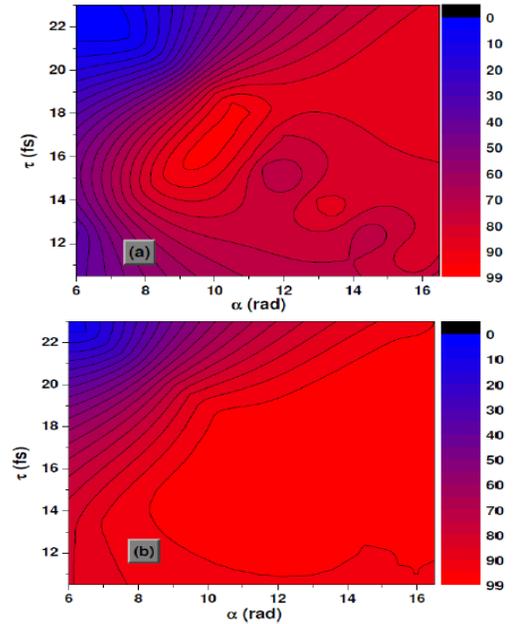

Fig. 3 (Color online) Contour plots of the final population (in %) (a) $\rho_{33}(\infty)$, (b) $\rho_{44}(\infty)$ for varying frequency

sweeping parameter, $\alpha$ and chirp steepening parameter, $\tau$, and other parameters are the same as those in Fig. 2.

A careful inspection of Fig. 3(a) reveals that final population in state $|3\rangle$, $\rho_{33}(\infty)$ is robust against the variation in the frequency sweeping parameter, $\alpha$ and chirp steepening parameter, $\tau$, to a sufficiently large range, e.g., $\alpha \approx$ 9-11 rad and $\tau \approx$ 15-18 fs respectively, which amount to more than 95 % population. On the other hand, it could be observed from Fig. 3(b) that final population in state $|4\rangle$, $\rho_{44}(\infty)$ is highly robust against the variation in same parameters, $\alpha$ and $\tau$, to a large range, $\alpha \approx$ 9-16 rad and $\tau \approx$ 12-20 fs respectively, which amount to more than 95 % population. The final population is found to be robust against the variation of other pulse parameters as well. For example, the final population $\rho_{33}(\infty)$ is robust against the variation of peak Rabi frequency in the range, 0.56 rad/fs - 0.63 rad/fs with more than 95 % populations and $\rho_{44}(\infty)$ is robust against the variation of peak Rabi frequency in the range, 0.50 rad/fs - 0.70 rad/fs with more than 95 % population. In addition, $\rho_{33}(\infty)$ and $\rho_{44}(\infty)$ are found to be nearly 96 % and 97 % for $\beta =\gamma =1$ respectively and nearly 92 % for $\beta =1.1$ and $\gamma =0.9$. However, one can achieve the more than 92 % population with $\beta =1.1$ and $\gamma =0.9$ by judiciously choosing the pulse parameters such as $\Omega_R(0)$, $\alpha$ and $\tau$ etc. For example, nearly 97 % population transfers to state $|3\rangle$ could be achieved with $\Omega_R(0) =0.55$ rad/fs, $\alpha =11.50$ rad and $\tau = 18$ fs.

In conclusion, we have demonstrated the ultrafast and selective population transfer in Y-like four level atoms by utilizing only one single frequency chirped few-cycle pulse. The selectivity is obtained by just manipulating the chirp offset parameter. The selective population transfer is found to be robust against the variation of simulation parameters and hence the scheme may be explored in the other atoms as well which can be modeled as Y-like four level atoms. The scheme may be explored in complex molecules as well owing the selectivity offered by the chirp offset parameter.